\documentclass[aps,showpacs,manuscript,12pt]{revtex4}
\usepackage{amssymb}
\usepackage{amsmath}
\usepackage{graphicx}

\begin{document}
\title{\textbf{Mathematical properties of the Navier-Stokes dynamical
system for incompressible Newtonian fluids$^{\S }$ }}
\author{Massimo Tessarotto$^{a,b}$, Claudio Asci$^{a}$, Claudio Cremaschini$^{c,d}$%
, Alessandro Soranzo$^{a}$ and Gino Tironi$^{a,b}$} \affiliation{\
$^{a}$Department of Mathematics and Informatics, University of
Trieste, Italy\\ $^{b}$Consortium for
Magneto-fluid-dynamics\thanks{Web site:
http://cmfd.univ.trieste.it}, University of Trieste, Italy\\
$^{c}$International School for Advanced Studies, SISSA, Trieste,
Italy \\ $^{d}$INFN, Trieste Section, Trieste, Italy }
\begin{abstract}
A remarkable feature of fluid dynamics is its relationship with
classical dynamics and statistical mechanics. This has motivated
in the past mathematical investigations concerning, in a special
way, the "derivation" based on kinetic theory, and in particular
the Boltzmann equation, of the incompressible Navier-Stokes
equations (INSE). However, the connection determined in this way
is usually merely asymptotic (i.e., it can be reached only for
suitable limit functions) and therefore presents difficulties of
its own. This feature has suggested the search of an alternative
approach, based on the construction of a suitable inverse kinetic
theory (IKT; Tessarotto et al., 2004-2008), which can avoid them.
IKT, in fact, permits to achieve an exact representation of the
fluid equations by identifying them with appropriate moment
equations of a suitable (inverse) kinetic equation. The latter can
be identified with a Liouville equation advancing in time a
phase-space probability density function (PDF), in terms of which
the complete set of fluid fields (prescribing the state of the
fluid) are determined. In this paper we intend to investigate the
mathematical properties of the underlying
\textit{finite-dimensional }phase-space classical dynamical
system, denoted \textit{Navier-Stokes dynamical system}, which can
be established in this way. The result we intend to establish has
fundamental implications both for the mathematical investigation
of Navier-Stokes equations as well as for diverse consequences and
applications in fluid dynamics and applied sciences.

\end{abstract}
\date{\today }
\maketitle


\section{1 - Introduction}

A fundamental theoretical issue in mathematical physics is the
search of a possible \textit{finite-dimensional} classical
dynamical system - to be denoted as \textit{Navier-Stokes} (NS)
\textit{dynamical system} - which uniquely advances in time the
complete set of fluid fields which characterize a fluid system. In
the case of an incompressible isothermal Newtonian fluid (also
known as \textit{NS fluid}) these are identified with the (mass)
fluid velocity and the (non-negative) scalar fluid pressure, which
in the domain of the fluid itself satisfy the
\textit{incompressible NS equations} (INSE). The reason why the
determination of such a dynamical system is so important is that
its existence is actually instrumental for the establishment of
theorems of existence and uniqueness for the related
initial-boundary value problem (\textit{INSE problem}). In this
paper we prove that, based on the inverse kinetic theory approach
(IKT) developed by
Tessarotto \textit{et al.} (2004-2007 \cite%
{Tessarotto2004,Ellero2005,Tessarotto2006}), the problem can
actually be given a well-defined formulation. Main goal of the
paper is the establishment of an \textit{equivalence theorem}
between the INSE problem and the NS dynamical system. Basic
consequences and applications of this result are pointed out. In
particular, contrary to the widespread view according to which the
phase-space dynamical system characterizing the fluid fields
should be infinite dimensional, here we intend to prove that a
\textit{finite-dimensional classical dynamical system} exists
which advances in time the complete set of fluid fields. This is
realized by the NS dynamical system.

\section{2 - The strong stochastic INSE problem}

For definiteness, let us{\ assume that the complete set of fluid
fields
fluid fields }$\left\{ Z\right\} \equiv \left\{ \rho ,\mathbf{V},p\right\} ,$%
{\ respectively denoting the mass density, the fluid velocity and
the fluid pressure, describing an NS fluid, are local strong
solutions of the equations }
\begin{eqnarray}
\rho &=&\rho _{o},  \label{1b} \\
\nabla \cdot \mathbf{V} &=&0,  \label{1ba} \\
N\mathbf{V} &=&0,  \label{1bb} \\
Z(\mathbf{r,}t_{o}) &\mathbf{=}&Z_{o}(\mathbf{r}),  \label{1ca} \\
\left. Z(\mathbf{r,}t)\right\vert _{\partial \Omega }
&\mathbf{=}&\left. Z_{w}(\mathbf{r,}t)\right\vert _{\partial
\Omega },  \label{1c}
\end{eqnarray}%
{In particular}, Eqs. (\ref{1b})-(\ref{1c}) are respectively the \textit{%
incompressibility, isochoricity and Navier-Stokes equations} and
the initial and Dirichlet boundary conditions for $\left\{
Z\right\} ,$ with $\left\{
Z_{o}(\mathbf{r})\right\} $ and $\left\{ \left. Z_{w}(\mathbf{\cdot ,}%
t)\right\vert _{\partial \Omega }\right\} $ suitably prescribed
initial and
boundary-value fluid fields, defined respectively at the initial time $%
t=t_{o}$ and on the boundary $\partial \Omega .$ {In the
remainder, for definiteness, we shall require that:}

\begin{enumerate}
\item {$\Omega ${\ is coincides with the Euclidean space }$E^{3}$ on }$%
\mathbb{R}
${$^{3}${\ (\textit{external domain}) and }}$\partial \Omega $
with the
improper plane of $%
\mathbb{R}
${$^{3};$}

\item $I$ is identified with the real axis $%
\mathbb{R}
$ (\textit{global domain}) .{\ }
\end{enumerate}

We shall assume that the fluid fields are continuous in $\overline{\Omega }%
\times {I}$ and fulfill the inequalities%
\begin{eqnarray}
&&\left. p>0,\right.  \label{5aa} \\
&&\left. \rho >0.\right.  \label{6aa}
\end{eqnarray}%
Here the notation as follows. $N$ is the \textit{NS nonlinear
operator}
\begin{equation}
N\mathbf{V}=\frac{D}{Dt}\mathbf{V}-\mathbf{F}_{H},  \label{NS
operator}
\end{equation}%
with $\frac{D}{Dt}\mathbf{V}$ and\textbf{\ }$\mathbf{F}_{H}$
denoting
respectively the \textit{Lagrangian fluid acceleration} and the \textit{%
total force} \textit{per unit mass}
\begin{eqnarray}
&&\left. \frac{D}{Dt}\mathbf{V}=\frac{\partial }{\partial t}\mathbf{V}+%
\mathbf{V}\cdot \nabla \mathbf{V,}\right.  \label{2a} \\
&&\left. \mathbf{F}_{H}\equiv \mathbf{-}\frac{1}{\rho _{o}}\nabla p+\frac{1}{%
\rho _{o}}\mathbf{f}+\upsilon \nabla ^{2}\mathbf{V,}\right.
\label{2c}
\end{eqnarray}%
while $\rho _{o}$ and $\nu >0$ are the constant\textit{\ mass
density} and the constant \textit{kinematic viscosity}. In
particular, $\mathbf{f}$ is the \textit{volume force density}
acting on the fluid, namely which is
assumed of the form%
\begin{equation}
\mathbf{f=-\nabla }\phi (\mathbf{r},t)+\mathbf{f}_{R},
\end{equation}%
$\phi (\mathbf{r},t)$ being a suitable scalar potential, so that
the first
two force terms [in Eq.(\ref{2c})] can be represented as $-\nabla p+\mathbf{f%
}$ $=-\nabla p_{r}+\mathbf{f}_{R},$ with
\begin{equation}
p_{r}(\mathbf{r},t)=p(\mathbf{r},t)-\phi (\mathbf{r},t),
\end{equation}%
denoting the \textit{reduced fluid pressure}. As a consequence of Eqs.(\ref%
{1b}) and (\ref{1ba}) it follows that the fluid pressure
necessarily
satisfies the \textit{Poisson equation}%
\begin{equation}
\nabla ^{2}p=S,  \label{Poisson}
\end{equation}%
where the source term $S$ reads
\begin{equation}
S=-\rho _{o}\nabla \cdot \left( \mathbf{V}\cdot \nabla
\mathbf{V}\right) +\nabla \cdot \mathbf{f}.
\end{equation}%
Here we shall assume, furthermore, that the fluid fields $\left\{
Z\right\} , $ together with the volume force density $\mathbf{f}$
and the initial and boundary conditions $\left\{
Z_{o}(\mathbf{r})\right\} ,$ $\left\{ \left.
Z_{w}(\mathbf{r,}t)\right\vert _{\partial \Omega }\right\} ${\ are
all
stochastic functions (see Appendix) of the form }%
\begin{eqnarray}
&&\left. Z=Z(\mathbf{r},t,\mathbf{\alpha }),\right.  \notag \\
&&\left. \mathbf{f}\mathbf{=f}(\mathbf{r},t,\mathbf{\alpha
})\right.  \notag
\\
&&\left. Z_{o}=Z_{o}(\mathbf{r},\mathbf{\alpha })\right. \\
&&\left. Z_{w}=\left. Z_{w}(\mathbf{r,}t)\right\vert _{\partial
\Omega }\right.  \notag
\end{eqnarray}%
where $\mathbf{\alpha }\in V_{\mathbf{\alpha }}$ are stochastic
variables assumed independent of $(\mathbf{r},t$).
Eqs.(\ref{1b})-(\ref{1c}) then denote the{\ initial-boundary value
problem for the }stochastic incompressible Navier-Stokes
equations{\ (\textit{strong stochastic INSE problem}). }

\section{3 - The IKT-statistical model}

A fundamental aspect of fluid dynamics lays in the construction of
statistical models for the fluid equations \cite{Tessarotto2009c}.\textit{\ }%
In this connection a possible viewpoint is represented by the
construction of the so-called \textit{IKT-statistical
model}\textit{\ }able to yield as moments of the PDF the
\textit{whole set of \ fluid fields }$\left\{ Z\right\} $ which
determine the fluid state and in which the same PDF satisfies a
Liouville equation.\textit{\ }Despite previous attempts (Vishik
and Fursikov, 1988 \cite{Vishik1988} and Ruelle, 1989
\cite{Ruelle1989}) the existence of such a dynamical system has
remained for a long time an unsolved problem.

This type of approach has actually been achieved for
incompressible NS fluids \cite{Tessarotto2004}. Its applications\
and extensions are
wide-ranging and concern in particular: incompressible thermofluids \cite%
{Tessarotto2008-2}, quantum hydrodynamic equations (see \cite%
{Tessarotto2007a,Tessarotto2008-4}), phase-space Lagrangian dynamics \cite%
{Tessarotto2008-5}, tracer-particle dynamics for thermofluids \cite%
{Tessarotto2008-6,Tessarotto2009b}, the evolution of the fluid
pressure in incompressible fluids \cite{Tessarotto2008-3},
turbulence theory in Navier-Stokes fluids
\cite{Tessarotto2008-4,Tessarotto2008-7} and magnetofluids
\cite{Tessarotto2009} and applications of IKT to lattice-Boltzmann
methods \cite{Tessarotto2008-8}.

The IKT-statistical model is based on the introduction of a PDF
depending on
the local configuration-space vector $\mathbf{r,}$ $f_{1}(t)\equiv f_{1}(%
\mathbf{r,v,}t,\mathbf{\alpha })$ (\textit{1-point velocity PDF})
defined on the phase-space $\Gamma _{1}=\Omega \times U$ [with
$U\equiv
\mathbb{R}
^{3}$] and identified with a strictly positive function such that
the complete set of fluid fields $\left\{ Z\right\} $ associate to
the strong
stochastic INSE problem can be represented in terms of the functionals (%
\textit{velocity moments})
\begin{equation}
\int\limits_{U}d\mathbf{v}Gf_{1}(t)=\left\{ 1,\mathbf{V}(\mathbf{r,}t,%
\mathbf{\alpha }),p_{1}(\mathbf{r,}t,\mathbf{\alpha })\right\} ,
\label{MOMENTS-2}
\end{equation}%
(\textit{Requirement \#1; correspondence principle}). Here $G=\left\{ 1,%
\mathbf{v,}u^{2}/2\right\} $while
$p_{1}(\mathbf{r,}t,\mathbf{\alpha })>0$ denotes the
\textit{kinetic pressure}
\begin{equation}
p_{1}(\mathbf{r},t,\mathbf{\alpha })=p(\mathbf{r},t,\mathbf{\alpha }%
)+p_{0}(t,\mathbf{\alpha })-\phi (\mathbf{r},t,\mathbf{\alpha }),
\label{kinetic pressure}
\end{equation}%
with $p_{0}(t,\mathbf{\alpha })>0$ (the \textit{pseudo-pressure})
a strictly
positive, smooth, i.e., at least $C^{(1)}(I)$, real function and $\phi (%
\mathbf{r},t,\mathbf{\alpha })$ a suitably defined potential. In addition $%
f_{1}(t)$ is assumed to obey the Liouville equation - or inverse
kinetic
equation (IKE) according to the notation of Ref.\cite{Ellero2005}) -%
\begin{equation}
L(\mathbf{r,v},t;f_{1}(t))f_{1}(t)=0  \label{LIOUVILLE EQ}
\end{equation}%
(\textit{Requirement \#2}) with $L(\mathbf{r,v},t;f_{1}(t))$
denoting the
Liouville streaming operator $L(\mathbf{r,v},t;f_{1}(t))\cdot \equiv \frac{%
\partial }{\partial t}\cdot +\frac{\partial }{\partial \mathbf{x}}\cdot
\left\{ \mathbf{X}(\mathbf{x},t;f_{1}(t))\cdot \right\} $ and $\mathbf{F}%
(f_{1}(t))$ a suitable smooth vector field defined in such a way
that the
moment equations of (\ref{LIOUVILLE EQ}) obtained for $G=\left\{ 1,\mathbf{%
v,\rho }_{o}u^{2}/3\right\} $ [$\mathbf{u}$ denoting the relative velocity $%
\mathbf{u\equiv u}(\mathbf{r,}t,\mathbf{\alpha })=\mathbf{v-V(r},t,\mathbf{%
\alpha })$] coincide respectively with Eqs.(\ref{1ba}),
(\ref{1bb}) and again (\ref{1ba}).\textit{\ }This implies that the
initial value problem
associated to the vector field%
\begin{equation}
\mathbf{X}(\mathbf{x},t;f_{1}(t))=\left\{
\mathbf{v,F}(f_{1}(t))\right\} , \label{vector field}
\end{equation}%
\begin{equation}
\left\{
\begin{array}{c}
\frac{d\mathbf{x}}{dt}=\mathbf{X}(\mathbf{x},t;f_{1}(t)), \\
\mathbf{x}(t_{o})=\mathbf{x}_{o}%
\end{array}%
\right.  \label{Eq.1}
\end{equation}%
necessarily defines a dynamical system. In particular if $\mathbf{x}(t)=%
\mathbf{\chi (x}_{o},t_{o},t\mathbf{)}$ is the general solution of (\ref%
{Eq.1}), this is identified with the flow%
\begin{equation}
T_{t_{o},t}:\mathbf{x}_{o}\rightarrow \mathbf{x}(t)=T_{t_{o},t}\mathbf{x}%
_{o}\equiv \mathbf{\chi (x}_{o},t_{o},t\mathbf{)}  \label{FLOW}
\end{equation}%
\textit{generated by }$\mathbf{X}(\mathbf{x},t;f_{1}(t))$\textit{. }%
Furthermore it is assumed that Eq.(\ref{LIOUVILLE EQ}) admits as a
particular solution $f_{1}(t)$ the Gaussian PDF%
\begin{equation}
f_{M}(\mathbf{x},t,\mathbf{\alpha })=\frac{1}{\pi ^{3/2}v_{th}^{3}(\mathbf{r,%
}t,\mathbf{\alpha })}\exp \left\{ -\frac{u^{2}}{v_{th}^{2}(\mathbf{r,}t,%
\mathbf{\alpha })}\right\} ,  \label{GAUSSIAN}
\end{equation}%
where $v_{th}^{2}(\mathbf{r,}t,\mathbf{\alpha })=2p_{1}(\mathbf{r,}t,\mathbf{%
\alpha })/\rho _{o}.$ More precisely it is assumed that $f_{M}(\mathbf{x},t,%
\mathbf{\alpha })$ is a particular solution of Eq.(\ref{LIOUVILLE
EQ}) if and only if the fluid fields
$Z(\mathbf{r},t,\mathbf{\alpha })$ are solutions of the strong
stochastic INSE problem (\textit{Requirement \#3}).

In the following we intend to investigate, in particular, the
consequences of requirements (\ref{MOMENTS-2}), (\ref{LIOUVILLE
EQ}) and (\ref{GAUSSIAN}) for the problem posed in this paper.

\section{4 - The Navier-Stokes dynamical system}

According to a certain misconception, dynamical systems for
continuous fluids cannot be finite dimensional due to the fact
that the fluid equations are PDEs for the relevant set of fluid
fields $\left\{ Z\right\} $. However, it is easy to show that this
is not the case even in the so-called Lagrangian description of
fluid dynamics. For a NS fluid this is realized by parametrizing
the fluid velocity $\mathbf{V}$ in terms of the Lagrangian path
(LP) $\mathbf{r}(t).$ In the present notation this is solution of
the problem
\begin{equation}
\left\{
\begin{array}{l}
\frac{D\mathbf{r}(t)}{Dt}=\mathbf{V}(\mathbf{r}(t),t,\mathbf{\alpha }), \\
\mathbf{r}(t_{o})=\mathbf{r}_{o},%
\end{array}%
\right.  \label{LP}
\end{equation}%
where $\frac{D}{Dt}$ is the Lagrangian derivative (\ref{2a}). As a
consequence the NS equation becomes
\begin{equation}
\frac{D\mathbf{V}(\mathbf{r}(t)\mathbf{,}t)}{Dt}=\mathbf{F}_{H}(\mathbf{r}(t)%
\mathbf{,}t),
\end{equation}%
which [with $\mathbf{F}_{H}(\mathbf{r}(t)\mathbf{,}t)$ considered
prescribed] can be treated as an ODE and integrated along a LP
yielding
\begin{equation}
\mathbf{V}(\mathbf{r}(t)\mathbf{,}t)=\mathbf{V}(\mathbf{r}_{o}\mathbf{,}%
t_{o})+\int\limits_{t_{o}}^{t}dt^{\prime }\mathbf{F}_{H}(\mathbf{r}%
(t^{\prime })\mathbf{,}t^{\prime }).
\end{equation}%
This permits to determine, \textit{for all} $\mathbf{r\equiv r}(t)$ \textit{%
and }$\left( \mathbf{r}_{o}\mathbf{,}t_{o}\right) \in \Omega
\times I,$ the
vector field $\mathbf{V}(\mathbf{r,}t)$. Therefore the \textit{%
finite-dimensional dynamical system} $\left( \mathbf{r}_{o}\mathbf{,}%
t_{o}\right) \rightarrow $\ $(\mathbf{r}(t),t)$ defined by
Eq.(\ref{LP})
actually generates the time evolution of $\mathbf{V}(\mathbf{r},t,\mathbf{%
\alpha }).$ Nevertheless, in this description the fluid pressure
is actually
not directly determined [in fact this requires solving the Poisson equation (%
\ref{Poisson})].

We intend to show that a dynamical system advancing in time the
complete set
of fluid fields for a NS fluid is realized by the dynamical system $%
T_{t_{o},t}$ [NS dynamical system; see Eq.(\ref{FLOW})]. Its
precise definition depends manifestly on the vector field
$\mathbf{F}(f_{1}(t))$. The task [of defining
$\mathbf{F}(f_{1}(t))]$ is achieved by the IKT approach developed
in Ref.\cite{Ellero2005}. As a consequence it follows
that $\mathbf{F}(f_{1}(t))$ can be non-uniquely determined \cite%
{Ellero2005,Tessarotto2006} in terms of a smooth vector field
which is at least $C^{(1)}(\Gamma _{1}\times I\times
V_{\mathbf{\alpha }})$.

In this case, as a fundamental mathematical result, we intend to
prove here \textit{the equivalence between the strong stochastic
INSE problem and the NS dynamical system. } In other words the
Liouville equation (\ref{LIOUVILLE EQ}) can be shown to be
equivalent to
\begin{equation}
f_{1}(\mathbf{x},t,\mathbf{\alpha })=J(t,\mathbf{\alpha
})f_{1}(\mathbf{\chi (x},t,t_{o},\mathbf{\alpha
)},t_{o},\mathbf{\alpha }), \label{INTEGRAL LIOUVILLE EQ.}
\end{equation}%
where $\mathbf{x}(t)=T_{t_{o},t}\mathbf{x}_{o}$ is the general solution of (%
\ref{Eq.1}) for $\mathbf{F}\equiv \mathbf{F}(f_{1}(t))$ and where%
\begin{equation}
J(t,\mathbf{\alpha })=\exp \left\{ \int\limits_{t_{o}}^{t}dt^{\prime }\frac{%
\partial }{\partial \mathbf{v(t}^{\prime },\mathbf{\alpha })}\cdot \mathbf{F}%
(\mathbf{x}(t^{\prime },\mathbf{\alpha }),t^{\prime },\mathbf{\alpha ;}%
f_{1}(t^{\prime }))\right\} \equiv \left\vert \frac{\partial \mathbf{x}(t,%
\mathbf{\alpha })}{\partial \mathbf{x}_{o}}\right\vert
\end{equation}
is the Jacobian of the flow $T_{t_{o},t}$ [see Eq.(\ref{FLOW})].
Therefore,
the NS dynamical system necessarily advances in time the PDF $f_{M}(\mathbf{x%
},t_{o},\mathbf{\alpha })$ so that it is \textit{identically} a
solution of the Liouville equation (\ref{LIOUVILLE EQ}). The
result\textit{\ }can be
established on general grounds, i.e., for an arbitrary vector field $\mathbf{%
F}((f_{1}(t))$ fulfilling Requirements \#1-\#3. The following result holds:%
\newline
\newline

\textbf{THM.1 - Equivalence theorem }

{\it In validity of Requirements $\#1-\#3$ the strong stochastic
INSE problem is equivalent to the NS dynamical system (\ref{FLOW})
}.

PROOF

The proof is immediate. In fact, if
\begin{equation}
\mathbf{x\equiv x}(t,\mathbf{\alpha })=\mathbf{\chi (x}_{o},t_{o},t,\mathbf{%
\alpha )}
\end{equation}%
is the solution of Eq.(\ref{Eq.1}) (which is assumed to exist and
define at least a $C^{(2)}-$diffeomorphism,\ its inverse
transformation is simply
\begin{equation}
\mathbf{x}_{o}=\mathbf{\chi (x},t,t_{o},\mathbf{\alpha ).}
\end{equation}%
Therefore by differentiating Eq.(\ref{INTEGRAL LIOUVILLE EQ.}) it follows%
\begin{equation}
\frac{d}{dt}f_{1}(\mathbf{x},t,\mathbf{\alpha })-\frac{d}{dt}\left\{ J(t,%
\mathbf{\alpha })f_{1}(\mathbf{\chi (x},t,t_{o},\mathbf{\alpha )},t_{o},%
\mathbf{\alpha })\right\} =0,
\end{equation}%
which recovers Eq.(\ref{INTEGRAL LIOUVILLE EQ.}) and admits as a
particular solution $f_{M}(t)\equiv
f_{M}(\mathbf{x},t,\mathbf{\alpha })$ when subject to the initial
condition $f_{M}(\mathbf{x}_{o},t_{o},\mathbf{\alpha })$. Hence in
terms of such an equation the NS dynamical system advances in time
\textit{the complete set of fluid fields. }Therefore, the fluid
velocity and
the kinetic pressure at time $t$, i.e.,$\mathbf{V}(t)\equiv \mathbf{V}(%
\mathbf{r},t,\mathbf{\alpha })$ and $p_{1}(t)\equiv p_{1}(\mathbf{r},t,%
\mathbf{\alpha }),$ follow from the moment equations
(\ref{MOMENTS-2}). Q.E.D.

\section{5. Conclusions}

This work is motivated by the analogy between hydrodynamic
description and the theory of classical dynamical systems. For
greater generality the case of stochastic fluid equations has been
considered. The problem of the equivalence between the
initial-boundary value problem for incompressible Navier-Stokes
equations and the Navier-Stokes dynamical system introduced in
Ref.\cite{Ellero2005} has been investigated. Indeed, the theory
here developed applies both to deterministic and stochastic fluid
fields. In fact, in both cases the time evolution of $f_{1}$ is
determined by a Liouville equation [see Eq.(\ref{LIOUVILLE EQ})]
which evolves in time also the complete set of fluid fields (all
represented in terms of moments of the same PDF). Contrary to the
misconception according to which the phase-space dynamical system
characterizing the fluid fields $\left\{ Z\right\} $ of a
continuous fluid system should be infinite dimensional, here we
have proven that the
finite-dimensional NS classical dynamical system advances in time \textit{%
the complete set of fluid fields,} determined in terms of velocity
moments of the 1-point PDF $f_{1}(\mathbf{x},t,\mathbf{\alpha })$.
The theory here developed applies generally to stochastic fluid
equations. As shown elsewhere
\cite{Tessarotto2008-7,Tessarotto2009,Tessarotto2009c}, this
represents a convenient treatment for the statistical theory of
turbulence,
historically referred to the work of Kolmogorov (Kolmogorov, 1941 \cite%
{Kolmogorov1941}) and Hopf (Hopf, 1950/51 \cite{Hopf1950/51}).

The theory has important consequences which concern fundamental
aspects of fluid dynamics:

\begin{itemize}
\item determination of the NS dynamical system advancing in time the
complete set of fluid fields of a turbulent NS fluid
\cite{Tessarotto2009c};

\item construction of the initial conditions for the 1-point PDF $f_{1}$
\cite{Tessarotto2010c}$;$

\item determination of the time-evolution of passive scalar and tensor
fields \cite{Tessarotto2008-3};

\item construction of the exact equations of motion for ideal
tracer-particle dynamics in a turbulent NS fluids
\cite{Tessarotto2009b};

\item construction of multi-point PDFs for turbulent NS fluids \cite%
{Tessarotto2010e};

\item statistical treatment of homogeneous, isotropic and stationary
turbulence based on IKT \cite{Tessarotto2010d}.
\end{itemize}

\section{Appendix - \textbf{\ }Stochastic variables}

Let $(S,\Sigma ,P)$ be a probability space; a measurable function $\mathbf{%
\alpha :}S\longrightarrow V_{\mathbf{\alpha }}$, where $V_{\mathbf{\alpha }%
}\subseteq
\mathbb{R}
^{k}$, is called \textit{stochastic} (or \textit{random})
\textit{variable}.

A stochastic variable $\mathbf{\alpha }$\ is called \textit{continuous} if%
\textit{\ }it is endowed with a \textit{stochastic model} $\left\{ g_{%
\mathbf{\alpha }},V_{\mathbf{\alpha }}\right\} ,$\textit{\ }namely
a real
function\textit{\ }$g_{\mathbf{\alpha }}$ (called as \textit{stochastic PDF})%
\textit{\ }defined on the set $V_{\mathbf{\alpha }}$ and such
that:

1) $g_{\mathbf{\alpha }}$ is measurable, non-negative, and of the
form
\begin{equation}
g_{\mathbf{\alpha }}=g_{\mathbf{\alpha
}}(\mathbf{r},t,\mathbf{\cdot }); \label{stochastic PDF}
\end{equation}

2) if $A\subseteq V_{\mathbf{\alpha }}$ is an arbitrary Borelian subset of $%
V_{\mathbf{\alpha }}$ (written $A\in \mathcal{B}(V_{\mathbf{\alpha
}})$), the integral
\begin{equation}
P_{\mathbf{\alpha }}(A)=\int\limits_{A}d\mathbf{x}g_{\mathbf{\alpha }}(%
\mathbf{r},t,\mathbf{x})  \label{dist-of-alpha}
\end{equation}%
exists and is the probability that $\mathbf{\alpha \in }A$; in
particular, since $\mathbf{\alpha }\in V_{\mathbf{\alpha }}$,
$g_{\mathbf{\alpha }}$ admits the normalization
\begin{equation}
\int\limits_{V_{\mathbf{\alpha }}}d\mathbf{x}g_{\mathbf{\alpha }}(\mathbf{r}%
,t,\mathbf{x})=P_{\mathbf{\alpha }}(V_{\mathbf{\alpha }})=1.
\label{normalization}
\end{equation}

The set function $P_{\mathbf{\alpha }}:\mathcal{B}(V_{\mathbf{\alpha }%
})\rightarrow \lbrack 0,1]$ defined by (\ref{dist-of-alpha}) is a
probability measure and is called distribution (or law) of $\mathbf{\alpha }$%
. Consequently, if a function $f\mathbf{:}V_{\mathbf{\alpha }%
}\longrightarrow V_{f}\subseteq
\mathbb{R}
^{m}$ is measurable, $f$ is a stochastic variable too.

Finally define the \textit{stochastic-averaging operator
}$\left\langle
\cdot \right\rangle _{\mathbf{\alpha }}$(see also \cite{Tessarotto2009}) as%
\textit{\ }%
\begin{equation}
\left\langle f\right\rangle _{\mathbf{\alpha }}=\left\langle f(\mathbf{y}%
,\cdot )\right\rangle _{\mathbf{\alpha }}\equiv \int\limits_{V_{\mathbf{%
\alpha }}}d\mathbf{x}g_{\mathbf{\alpha }}(\mathbf{r},t,\mathbf{x})f(\mathbf{y%
},\mathbf{x}),  \label{stochastic averaging operator}
\end{equation}%
for any $P_{\mathbf{\alpha }}$-integrable function $f(\mathbf{y},\cdot ):V_{%
\mathbf{\alpha }}\rightarrow
\mathbb{R}
$, where the vector $\mathbf{y}$ is some parameter.

\section*{Acknowledgments}
Work developed in cooperation with the CMFD Team, Consortium for
Magneto-fluid-dynamics (Trieste University, Trieste, Italy).
Research partially performed in the framework of the GDRE (Groupe
de Recherche Europ\'{e}en) GAMAS.



\end{document}